\documentclass[rnote]{aa} % for the research notes
\usepackage{graphicx}
\usepackage{natbib}
\usepackage{txfonts}
\usepackage{array}
\usepackage{rotating}
\def\cyg{Cyg\,OB2\,\#9}
\def\l{$\lambda$}
\def\ha{H$\alpha$}

\def\hei{He\,{\sc i}}
\def\heii{He\,{\sc ii}}

\def\ciii{C\,{\sc{iii}}}
\def\civ{C\,{\sc{iv}}}
\def\oiii{O\,{\sc{iii}}}
\def\kms{km\,s$^{-1}$}

\begin{document}
   \title{A binary signature in the non-thermal radio-emitter \cyg\thanks{Based on observations obtained at the OHP and the Asiago observatory.}}

   \author{Ya\"el Naz\'e
          \inst{1}\fnmsep\thanks{Postdoctoral Researcher FRS/FNRS}
          \and
          Micha\"el De Becker\inst{1}\fnmsep$^{\star\star}$
          \and
          Gregor Rauw\inst{1}\fnmsep\thanks{Research Associate FRS/FNRS}
          \and
          Cesare Barbieri\inst{2}
          }

   \offprints{Y. Naz\'e}

   \institute{Institut d'Astrophysique et de G\'eophysique, Universit\'e de Li\`ege, All\'ee du 6 Ao\^ut 17 Bat. B5C, B4000-Li\`ege, Belgium\\
              \email{naze@astro.ulg.ac.be}
         \and
             Dipartimento di Astronomia, Universit\'a degli studi di Padova, vicolo Osservatorio 2, I35122-Padova, Italy
             }

   \date{Received January 31, 2008; accepted March 17, 2008}
 
  \abstract
   {} %leave it empty if necessary  
   {Non-thermal radio emission associated with massive stars is believed to arise from a wind-wind collision in a binary system. However, the evidence of binarity is still lacking in some cases, notably \cyg.}
   {For several years, we have been monitoring this heavily-reddened star from various observatories. This campaign allowed us to probe variations both on short and long timescales and constitutes the first in-depth study of the visible spectrum of this object. }
   {Our observations provide the very first direct evidence of a companion in \cyg, confirming the theoretical wind-wind collision scenario. These data suggest a highly eccentric orbit with a period of a few years, compatible with the 2yr-timescale measured in the radio range. In addition, the signature of the wind-wind collision is very likely reflected in the behaviour of some emission lines.}
   {}


   \maketitle
%
%________________________________________________________________

\section{Introduction}
The Cyg\,OB2 association constitutes one of the richest OB associations of our Galaxy. Using 2MASS data, \citet{kno00} estimated that it contains nearly 3000 hot stars, among them more than a hundred O-type stars. This cluster is therefore an ideal target for investigations of the massive star population. However, because of their high reddening, the nature of the main stars of Cyg\,OB2 only begins to be uncovered. 

   \begin{figure*}
   \centering
   \includegraphics[width=18cm, bb=18 320 592 718, clip]{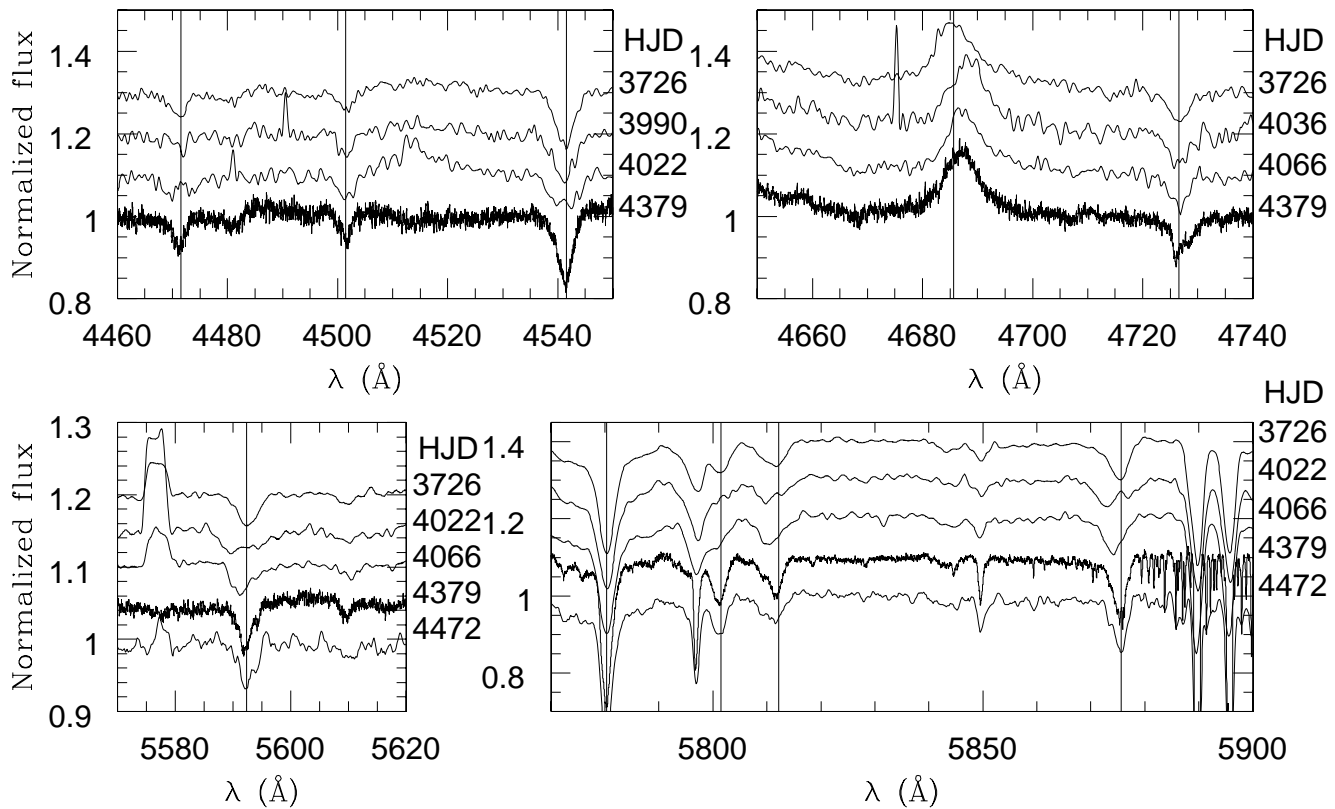}
   \caption{Evolution of the line profiles in a few selected spectra (dates shown to the right in format HJD-2\,450\,000). Vertical lines are drawn at the rest wavelength of the stellar lines and at the DIBs wavelength observed in the Sophie high-resolution spectra. Compare the profiles of Dec. 2005 (date=3726), Sep. 2006 (3990), Oct. 2007 (4379), and Jan. 2008 (4472) to the ones of Oct. 2006 (4022 or 4036) and Nov. 2006 (4066). The saturated emission line close to O\,{\sc{iii}}\,$\lambda$\,5592 is a mercury night sky line from light pollution.}
              \label{others}
    \end{figure*}

Among the most prominent O-type stars of Cyg\,OB2, there are three objects that are known to be non-thermal radio emitters: Cyg\,OB2\,\#5, \#8a, and \#9. Such non-thermal radio emission is believed to arise from synchrotron radiation, probably produced by relativistic electrons accelerated through the first order Fermi mechanism in a hydrodynamic shock \citep[see e.g.][for a recent quantitative model]{pit06}. For massive stars, the requested shock could either form following instabilities intrinsic to the line-driven stellar wind of a single massive star or be a consequence of the collision of two stellar winds in a massive binary. Theoretical considerations support only the latter interpretation \citep{van05}, but observational evidence was still lacking until recently. Thanks to a dedicated monitoring, the signature of a companion was finally unveiled in the non-thermal emitters Cyg\,OB2\,\#8a and 9\,Sgr \citep[see][ for a review see also \citealt{deb07}]{deb04,rau05}.

However, \cyg\ (=VI Cyg 9, Schulte 9, [MT91] 431) remained one of the most challenging objects in this respect. On the one hand, \citet{pig98} reported some short-term (a few days), low amplitude ($\Delta I$=0.03mag) variations, of unknown origin, in the photometry of \cyg. On the other hand, the X-ray spectrum of this object is compatible with a colliding-wind (CW) binary \citep{rau05b}, but no optical signature of a companion has ever been reported. This could be explained by a simple lack of data, which is why we undertook a dedicated monitoring of this system a few years ago. Our observing campaign was aimed at probing short-term variations, but also long-term ones because the radio lightcurve indicates a period of $\sim$2.4yrs \citep{van08}. 

This paper is organised as follows. Section 2 presents the datasets and their reduction; section 3 shows the results obtained while section 4 provides a short summary and perspectives.

\section{Observations and data reduction}

Observations were carried out at two different observatories: Asiago observatory (Italy) and Haute-Provence observatory (OHP, France). A journal of the observations is provided in Table \ref{journ}. 

In Asiago, spectra were taken in 2005 and 2006 using the AFOSC (Asiago Faint Object Spectrograph \& Camera) in echelle mode (grisms 9 \& 10). Exposures generally lasted 1200s and provided medium-resolution spectra. Twelve orders were extracted, corrected for the blaze using flat fields, and calibrated. Due to the low resolution, blends severely affected the ThAr spectra, rendering the wavelength calibration quite rough. Therefore, we used the Diffuse Interstellar Bands (DIBs) to further refine the calibration (see below). Normalization was done for each order by fitting low-order polynomials into carefully chosen continuum windows.

Archival data were also found in the Asiago archives. Three low-resolution spectra were taken with a Boller \& Chivens spectrograph in 1996 while 14 medium-resolution spectra were obtained in 2003 with an holographic grating (VPH4). The typical exposure times amounted to 600s and 2400s, respectively. These data were reduced in a standard way.

At the OHP, the Aur\'elie spectrograph equipped with grating \#3 provided 27 additional spectra in the interval 2003-2008 while the Sophie echelle instrument observed the system 8 times in the high-efficiency mode. For individual spectra, the typical exposure time was 1800--3600s, and the data were finally smoothed by a moving box average. 

After a first analysis, it appeared that the system underwent only long-term changes. Therefore, spectra taken within 1--15days were generally averaged (see number N in Table \ref{journ}). To improve the wavelength calibration, we took advantage of the high reddening and used several narrow, well-marked DIBs close to major spectral lines. Their mean radial velocity (RV) measured on Sophie spectra, our highest resolution data, was chosen as reference. RVs amount to --15.2, +8.6, --8.0, --11.3\kms\ for DIBs at 4501.7, 4726.4, 5780.45 and 6613.62\AA, respectively. The stellar lines were fitted by gaussian(s) and the measured RVs were corrected by the shift derived from the closest DIB. This ensures that the RVs are correct to within 10\kms; larger excursions of the RVs are thus very likely real.

   \begin{table}
      \caption{Journal of observations. Julian dates (mean values if N$\neq$1) are given in the format HJD--2\,450\,000, N is the number of spectra taken, $\Delta\lambda$ is the wavelength range, R is the spectral resolution ($\lambda/FWHM_{calib}$), S/N is the average signal-to-noise ratio of the individual spectra. }
         \label{journ}
     \centering
         \begin{tabular}{lccccc}
            \hline\hline
Instrument &  Date & N & $\Delta\lambda$ (\AA)& R & S/N \\
            \hline
B\&Ch.    &  287.56 & 1 & 4900-6050 & 1000 & 300\\
          &  287.57 & 1 & 5950-7100 & 1200 & 500\\
          &  287.55 & 1 & 3850-5000 &  800 &  70\\
AFOSC-ls  & 2811.50 & 3 & 6400-7000 & 3400 & 300\\
          & 2857.30 & 5 & 6400-7000 & 3400 & 400\\
          & 2956.37 & 2 & 6400-7000 & 3400 & 300\\
          & 2987.74 & 4 & 6400-7000 & 3400 & 300\\
AFOSC-ech & 3726.23 & 4 & 3700-8800 & 3600 & 300\\
          & 3887.49 & 1 & 3700-8800 & 3600 & 200\\
          & 3990.46 & 1 & 3700-8800 & 3600 & 130\\
          & 4022.25 & 3 & 3700-8800 & 3600 & 150\\
          & 4036.24 & 2 & 3700-8800 & 3600 & 130\\
          & 4051.39 & 1 & 3700-8800 & 3600 & 160\\
          & 4066.30 & 1 & 3700-8800 & 3600 & 120\\
Aur\'elie & 2920.30 &16 & 6400-6750 &11000 &  70\\
          & 3290.29 &10 & 6400-6750 &11000 & 150\\
          & 3551.59 &14 & 6400-6750 &11000 & 250\\
          & 3651.71 &10 & 6400-6750 &11000 & 130\\
          & 4244.48 & 2 & 5500-5900 & 8800 &  50\\
          & 4303.50 & 2 & 5500-5900 & 8800 &  70\\
          & 4324.43 & 2 & 5500-5900 & 8800 & 100\\
          & 4472.25 & 1 & 5500-5900 & 8800 &  90\\
Sophie    & 4348.77 & 3 & 3900-6900 & 35000&  80\\
          & 4379.06 & 3 & 3900-6900 & 35000&  90\\
          & 4463.75 & 2 & 3900-6900 & 35000& 100\\
            \hline
         \end{tabular}

   \end{table}

\section{Data Analysis}

The most striking feature of the spectrum of \cyg\ is the presence of strong interstellar lines due to the high absorption of the star. Even some lines which are usually negligible in massive stars' spectra are here clearly detected \citep[for a list of DIBs see][]{her95}. Some can be mistaken for stellar features and one must be particularly careful in the analysis when the DIBs are close to actual stellar lines. For example, \civ\,\l\,5812 presents a double-line profile in the recent Sophie data, but this is only due to the contamination by a DIB at 5809.13\AA ! In lower-resolution spectra, such blends may remain undetected but they can still affect the measure of RVs (e.g. for \heii\,\l\,5412 and \civ\,\l\l\,5801,5812).

The observed spectrum of \cyg\ also displays the absorption lines typical of an O5.5If star: strong \heii\,\l\,4542,5412,6683 and weaker \hei\ lines (at 4471, 5876\AA). It also shows some emission features, notably \heii\,\l\,4686, \ciii\,\l\,5696, \ha. 

The first spectra of the monitoring (until September 2006) showed relatively narrow and strong absorption lines, but the situation changed abruptly in October 2006. At that time, the lines became shallower and broader, even showing double line profiles, the typical signature of a binary. In November 2006, the  lines presented intermediate line profile while the most recent spectra again showed stronger and narrower lines. This evolution is illustrated in Figure \ref{others}. The double-peaked profiles of uncontaminated stellar lines  provide the first undisputable evidence of the binary nature of \cyg. This statement stands on solid ground: it relies on an homogeneous dataset, the Asiago spectra, and is thus independent of instrument changes; moreover, such double line profiles cannot be spontaneously generated in the sole stellar lines by instrumental problems or reduction processes.

\begin{figure}
   \centering
   \includegraphics[width=10cm]{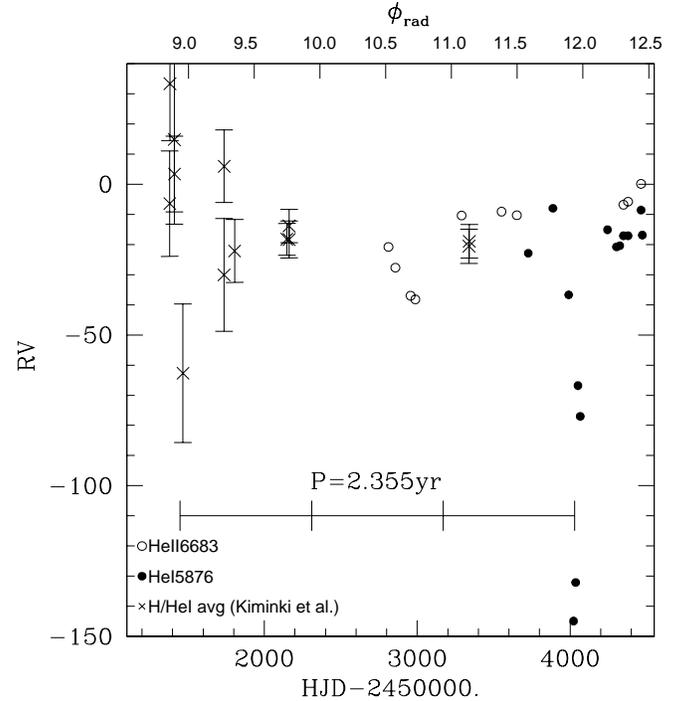}
   \caption{Radial velocity evolution of the best stellar lines (in terms of strength and non-contamination by DIBs) between June 2003 and January 2008. Average RV values for various H{\sc i}\ and \hei\ lines from \citet{kim07} are also presented. A line indicating the length of three periods of \citet{van08} is drawn for comparison and phases derived from their ephemeris are shown on top  (uncertainties amount to $\pm$0.08 in phase). }
              \label{rv}
    \end{figure}

A non-circular orbit was suggested by the variations of the radio emission \citep{van08}. The rapid variation of the line profiles in the visible domain also revealed the presence of eccentricity, with a quite high value. Figure \ref{rv} shows the evolution of the RVs with time for \hei\,\l\,5876 and \heii\,\l\,6683, two strong and uncontaminated stellar lines. A sharp change of the RVs is detected in \hei\,\l\,5876, as well as for other photospheric lines (\oiii\,\l\,5592 and \civ\,\l\l\,5801,5812). It is reminiscent of the periastron passage in a highly eccentric binary. In this context, it should be noted that the line splitting and large RV excursion occur when the radio emission is minimum ($\phi_{\rm Oct 06}\sim11.94\pm0.08$ following the ephemeris of \citealt{van08}), as expected if the non-thermal radio emission arises in the colliding-wing region.

The onset of a drop in the RVs can also be detected in the \heii\,\l\,6683 observations taken in 2003: the measured variations are thus compatible with the period of 2.355yr derived from the changes of the radio fluxes \citep{van08}, though the periastron passage itself unfortunately occurred during a gap in the observations. In this context, it should be noted that the average value of the RVs from \citet{kim07}, $-$16.1\kms, is fully compatible with our mean RV measured outside large RV excursion events and that the lowest value of the RV reported by these authors occured 3 periods before our 2006 RV drop (Fig. \ref{rv}). Unfortunately, as for us in 2003, \citet{kim07} missed the periastron passages themselves by only a week or two.

The splitting observed in October 2006 for \hei\,\l\,5876 allows us to derive RVs of +92 and --145\kms, whereas the RV of the apparently single component was --17\kms in October 2007. This suggests a mass ratio of about unity. The relative strength of the two components (see \hei\,\l\,5876 and \civ\,\l\,5812 profiles in Figure \ref{others}) suggests a slightly later type for the companion. The system could thus be O5+O6-7. However, the noise and uncertainties of our data prevent us from deriving more accurate spectral and orbital parameters.  

Finally, the non-thermal nature of the radio emission not only suggested a binary nature for \cyg, but also implies the presence of colliding winds. The signature of such a phenomenon can also be detected in the visible domain, as has been shown for several systems \citep[see e.g.][]{rau01,san01}. In this context, it is interesting to note that the RVs of the \heii\,\l\,4686 emission line do not follow the trend of the other lines: this line appears redshifted while the main component of the others lines are moving bluewards (Figure \ref{others}). Since the lines from the companion appear of slightly reduced strength compared to those of the primary star, it is possible that this line arises in the CW region. In addition, the \ha\ profile displays a complex behaviour. The absorption component first presents a rather triangular shape and then appears more rounded; the emission component strengthens when a maximum RV separation is seen in the photospheric lines -- such an anti-phase effect is quite typical of CW binaries.

   \begin{figure}
   \centering
   \includegraphics[width=8cm, bb=270 150 586 695, clip]{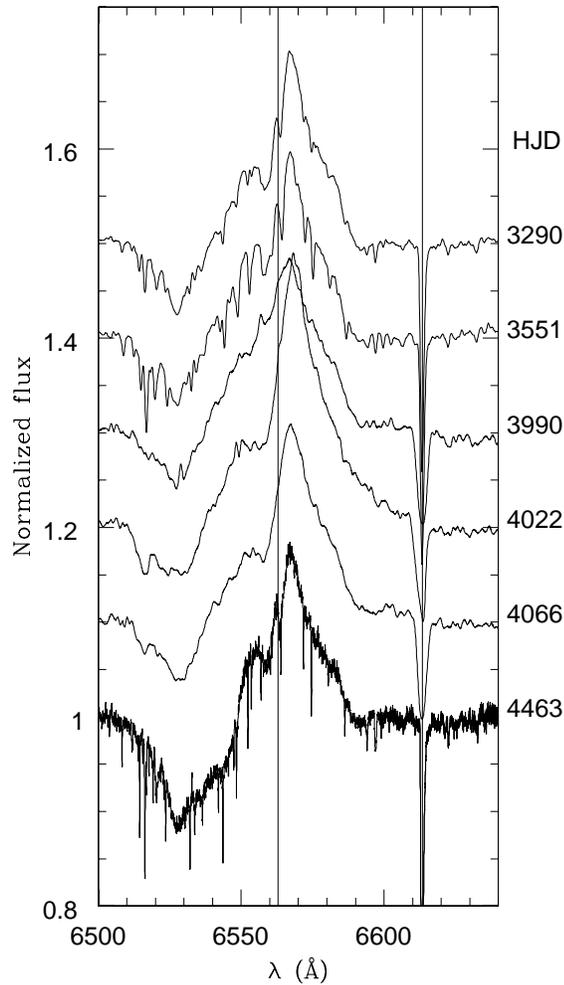}
   \caption{Evolution of the profile of the \ha\ line. The data were taken in Oct. 2004 (date=3290), June 2005 (3551), Sep. 2005 (3990), Oct. 2006 (4022), Nov. 2006 (4066) and Oct. 2007 (4379). Note the sharp increase in line strength at date 4022, i.e. when the absorption lines are showing their largest separation (see Fig. \ref{rv}).}
              \label{ha}
    \end{figure}

\section{Summary and conclusions}

A dedicated monitoring of \cyg\ has been going on since 2003. The detected line profile variations provide the first direct evidence of the presence of a companion in an eccentric orbit. The period of the detected changes is compatible with the 2.355yr timescale derived from radio measurements and the line splitting corresponds to the minimum emission in the radio range. The behaviour of the \heii\,\l\,4686 and \ha\ emission lines further suggests the presence of colliding winds. The binary status of \cyg\ lends thus additional support to the `standard scenario' for the non-thermal emission from early-type stars, where particle acceleration and synchrotron radio emission take place in the wind interaction region of a binary system \citep{deb07}. In addition, the quite high plasma temperature ($\sim$30 MK) derived from the fit of thermal models to the X-ray spectrum of \cyg\ is compatible with a scenario where a significant fraction of the X-rays are produced in a long period colliding wind binary \citep{rau05b}. 

Though the presented evidence for binarity is indisputable, much work is still needed to gain a complete knowledge of this peculiar massive system. To derive accurate orbital parameters and constrain the wind collision properties, it is necessary to accumulate more data. Additional spectra need to be taken with both high-resolution and high signal-to-noise, like e.g. our Sophie observations. It is particularly important to sample the rapid variations that occur near periastron. If October 2006 was indeed the last periastron event and if the 2.4yr radio period is correct, then the next periastron passage will take place in early 2009 - at a time when the star is not easily observable under good conditions (conjonction with the Sun). Nevertheless, any effort should be taken to monitor the system as close as possible to the event. A better estimate of the properties of \cyg\ might indeed require to wait for mid-2011, except if additional archival data, taken at the right epochs, are available.

\begin{acknowledgements}
We acknowledge support from the Fonds National de la Recherche Scientifique (FNRS, Belgium), the Scientific Cooperation program 2005-2006 between Italy and the Belgian `Communaut\'e Fran\c caise' (project 05.02), the OPTICON trans-national access programme, and the PRODEX XMM and Integral contracts (Belspo). 
\end{acknowledgements}

\end{document}